\documentclass[twocolumn,showpacs,preprintnumbers,amsmath,amssymb]{revtex4}

\usepackage{graphicx,color, amssymb}
\usepackage{dcolumn, wasysym}
\usepackage{bm}

\begin{document}

\preprint{APS/123-QED}

\title{ KFe$_{2}$As$_{2}$: coexistence of superconductivity and local moment
derived spin-glass
phases} 

\author{V.\ Grinenko}\affiliation{IFW Dresden, D-01171 Dresden, Germany}
\author{M.\ Abdel-Hafiez}\affiliation{IFW Dresden, D-01171 Dresden, Germany}
\author{S.\ Aswartham}\affiliation{IFW Dresden, D-01171 Dresden, Germany}
\author{A.U.B.\ Wolter-Giraud}\affiliation{IFW Dresden, D-01171 Dresden, Germany}
\author{C.\ Hess}\affiliation{IFW Dresden, D-01171 Dresden, Germany}
\author{M.~Kumar}\affiliation{IFW Dresden, D-01171 Dresden, Germany}
\author{S.\ Wurmehl}\affiliation{IFW Dresden, D-01171 Dresden, Germany}
\author{K.\ Nenkov}\affiliation{IFW Dresden, D-01171 Dresden, Germany}
\author{G.\ Fuchs}\affiliation{IFW Dresden, D-01171 Dresden, Germany}
\author{B.\ Holzapfel}\affiliation{IFW Dresden, D-01171 Dresden, Germany}
\author{S.-L.\ Drechsler}\affiliation{IFW Dresden, D-01171 Dresden, Germany}
\author{B. B\"uchner}\affiliation{IFW Dresden, D-01171 Dresden, Germany}

\date{\today}

\begin{abstract}
\noindent
High-quality KFe$_{2}$As$_{2}$ single crystals have been
studied by transport, magnetization and low-$T$ specific heat
measurements. Their
 analysis shows
that superconductivity occurs (in some cases coexists) 
in the vicinity of disordered magnetic phases (Griffiths
and spin glass type) 
depending of 
the amount of local magnetic moments
(probably excess Fe derived)
 and sample inhomogeneity.
The achieved phenomenological description
including also data from the literature
provides a consistent explanation of the observed non-Fermi-liquid 
behavior and of the {\it nominally} large 
experimental
Sommerfeld coefficient 
$\gamma_{\rm n} \approx 94$~mJ/mol$\cdotp$K$^2$. 
We suggest that the intrinsic value (directly related
to the itinerant quasi-particles)
$\gamma_{\rm el}\approx 
60(10)$~mJ/mol$\cdotp$K$^2$ is significantly reduced compared with 
$\gamma_{\rm n}$.  
Then an enhanced
$\Delta C_{el}/\gamma_{\rm el}T_c\sim$ 0.8 and  
a  weak total electron-boson coupling constant 
$\lambda \lesssim$ 1 follow.
    
\end{abstract}

\pacs{74.70.Xa, 74.62.En,
74.25.-q, 74.25Bt,  75.50}
\maketitle

The nature of the 
superconductivity in 
KFe$_{2}$As$_{2}$ (K122) is under debate 
\cite{Suzuki2011,Lee2011,Sato2009,Pascher2010,Zhang2010,Maiti2011,Thomale2011}.
This is due to its distinctive
characteristics with respect to other Fe-pnictides:
its heavily hole doping is 
 responsible for the lacking nesting
 of the Fermi surface
in contrast to less hole-doped
Ba$_{0.6}$K$_{0.4}$Fe$_{2}$As$_{2}$ \cite{Sato2009}. 
For instance, 
a neutron scattering study of K122
revealed well-defined low-energy incommensurate
spin fluctuations at $[\pi(1 \pm 2 \delta), 0]$ with $\delta$ =
$0.16$ \cite{Lee2011}. Also, NMR studies suggest that 
a type of AFM spin fluctuations {\it different} from 
that of the undoped 
(Ba, Ca, Sr)122 parent compounds develops in K122 \cite{Zhang2010}.
A strong or sizable effective mass enhancement of 
the quasi-particles has been observed in
de Haas-van 
Alphen \cite{Terashima2010, Hashimoto2010}  
and cyclotron resonance \cite{Kimata2011} measurements,  
respectively. Additionally, ARPES-data
also point to 
strong band renormalization which suggest an effective mass enhancement
by a factor of 2-4 \cite{Sato2009}.  The Sommerfeld coefficient 
$\gamma_{n} \sim$ 70 -100~mJ/mol\textperiodcentered K$^2$ from the linear 
specific heat (SH)
\cite{Fukazawa2011, Fukazawa2009, Kim2011} 
is strongly enhanced compared with the 
\begin{figure}[b]
\includegraphics[width=20pc,clip]{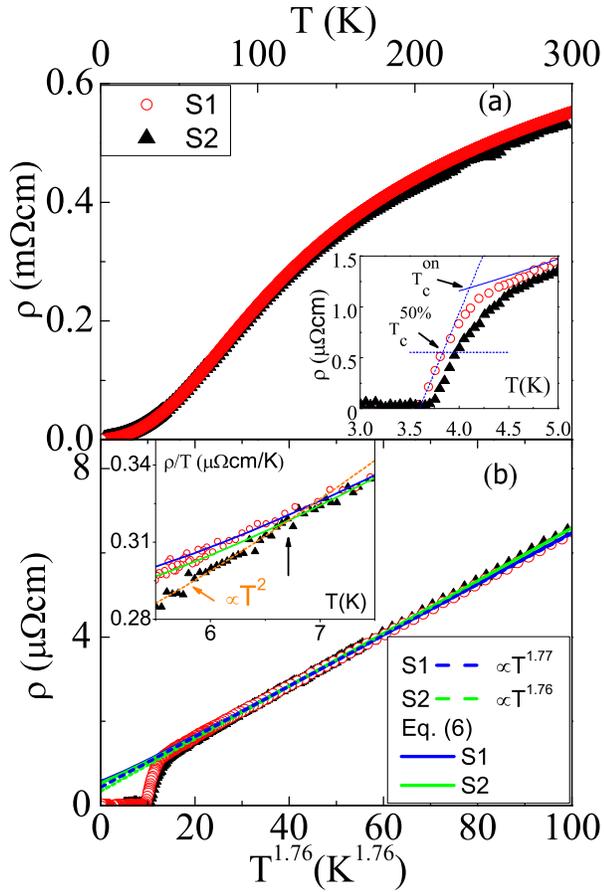}
\caption{(Color)(a) $T$-dependence of the in-plane resistivity $\rho$ of 
two K122 single crystals measured in zero-field up to $T \simeq $300~K. 
Inset: zoom into the superconducting state. (b) The
 $\rho$-data below 15~K
 plotted vs.\ $T^{1.5}$. Inset: 
$T$-dependence of the ratio $\rho$/T in the range $5.5\leq T \leq 7.5$~K
(see also text).}
\label{Fig:1}
\end{figure}
reported (from band structure density of states (DOS) derived)
"bare" values $\gamma_{\mbox{\tiny b}} \approx  10.1$ 
\cite{Terashima2010, Popovich2010} or 13 (mJ/mol$\textperiodcentered$K$^2$) 
\cite{Hashimoto2010} adopted from DFT-calculations. 
To the best of our knowledge, only for
one,
very clean, sample with
RRR$_5=\rho({\rm300\,K})/\rho ({\rm5\,K})
\approx 480$, a convincing $\rho \propto  T^2$ 
 was observed at low 
$T$ = 5 - 10~K evidencing 
standard Fermi-liquid behavior (FLB)
\cite{Hashimoto2010}. 
But for less perfect
samples with RRR$_5\approx 86$ 
the resistivity follows a 
{\it different, subsquared}
power-law:
$\rho \propto T^{1.5}$ above $T_c$  \cite{Dong2010},  
whose authors considered
this $T^{1.5}$-law 
as a 
signature of 
spin fluctuations in a
clean 3D-AFM. 
From the analysis of SH-data  
it was concluded that non-Fermi-liquid behavior (NFLB)
can be related to  magnetic
impurities \cite{Kim2011}. At low $T \ll T_c$, 
a further not yet fully understood 
magnetic anomaly has been observed. 
In this situation, the elucidation of the pairing
symmetry 
is a delicate problem and
a clear understanding of
various coexisting or competing forms of magnetism
should be addressed first. 

Similarly as
for La-1111 \cite{Grinenko2011}, Co and K doped 122 
\cite{Ni2008, Mukhopadhyay2009},
we suppose that 
point defects might induce 
local magnetic moments (LM)
also in K122. Recently, we have shown  
that LM 
in pnictides can be formed e.g.\ 
around As vacancies \cite{Grinenko2011}.
At variance with La-1111 
superconductors where 
such LM
only enhance the spin 
susceptibility \cite{Grinenko2011}, in K122 even 
a very small amount
of them leads to the formation of 
disordered magnetic phases such as spin glass (SG) and  Griffiths (G) phases. 
By analyzing carefully the low-$T$ behavior of $\rho(T)$, 
the magnetic susceptibility
$\chi(T)$, 
and SH data for K122 
single crystals with different amount of disorder,  
we will show that most of their
anomalous properties  
listed above are related to the vicinity and 
even to the
coexistence of 
superconductivity with these magnetic phases. 
This naturally explains the  NFLB of $\rho(T)$ and 
the unexpectedly high value of $\gamma_n$ obtained from 
low-$T$ SH data. 

K122 single crystals have been grown using a
self-flux method (FeAs-flux (S1) and KAs-flux (S2)), for description see 
\cite{Hafiez2011, Aswartham2012}.
Low-$T$ SH, ac susceptibility and four-probe resistivity were 
determined using a PPMS from Quantum Design.
The dc magnetic
susceptibility has  been measured 
 in a SQUID 
Quantum Design magnetometer.
Fig. \ref{Fig:1}(a) presents the $T$-dependence of the in-plane
electrical resistivity $\rho$ for both K122
single crystals S1 and S2. Upon cooling the $\rho$ decreases monotonically
showing a metallic behavior at all $T$ with a 
RRR$_5\approx$380 for crystal S1 and 
400 for S2. Below 10K the 
resistivity of both crystals shows NFLB:
$\rho(T)=\rho_0+AT^{\alpha}$ with $1.5 < \alpha  <2$
(Fig.\ \ref{Fig:1}(b)). 
Noteworthy, our $\alpha$ differs from
1.5 reported  for less perfect crystals
with RRR$_5 \approx 86$ 
\cite {Dong2010}, only, and
from 2 for the cleanest available case RRR$_5 \approx 480$ 
\cite{Hashimoto2010} 
where FLB has been  reported.
In this context we stress that the 
reported $T^2$-law for samples with
RRR$_5\sim 80$ and $T_c=2.8$~K, only, in  Ref.\ \onlinecite{TERASHIMA2009} might 
be related to a {\it 
too large} fit region up to 45~K used there.  
In fact, we observed that
too broad $T$-ranges can
{\it mask} deviations from the FLB \cite{remminorphases2}. 
Our samples have rather high transition temperatures:  
$T^{50\%}_{c}$=3.85(10)~K for 
sample S1 and 3.95(10)~K for sample S2 
(Fig.\ref{Fig:1}(a)) comparable with
the high
$T_c$-values of  
other single crystals with large 
RRR values \cite{Hashimoto2010, Kim2011, Fukazawa2011}. 
\begin{figure}
\includegraphics[width=20pc,clip]{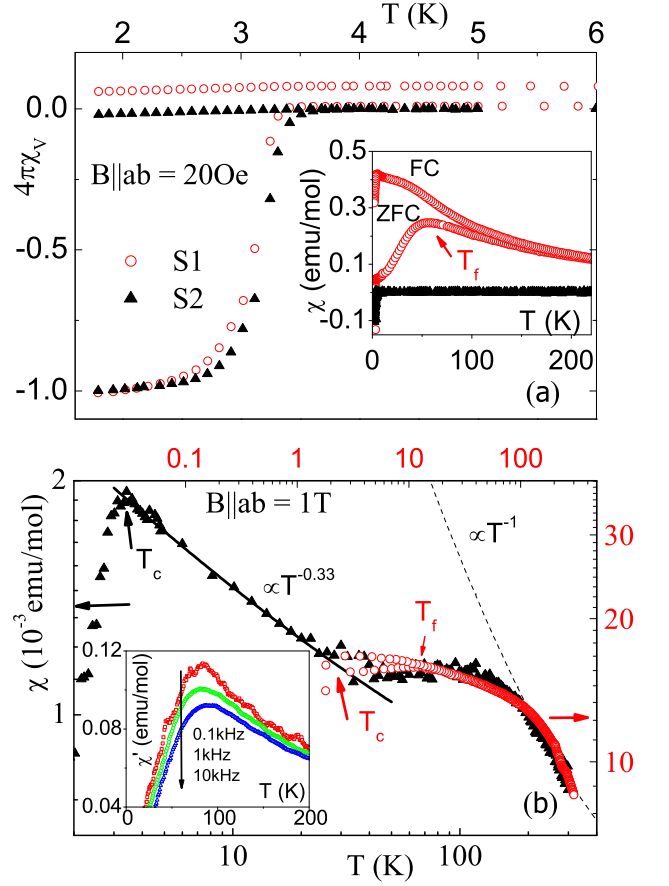}
\caption{(Color) (a) The $T$-dependence of the volume susceptibility
from dc magnetization of the two
K122 samples. Inset: $T$-dependence of the molar
susceptibility  for the same samples.
(b) $T$-dependence of the molar susceptibility at
$B \parallel ab$ = 1\,T (note the different axes 
for S1 and S2). For the interpretation of the fits, see text. 
Inset: $T$-dependence of the real part of
the ac susceptibility for various
frequencies of sample S1. The measurements
were done at 5~Oe ac field amplitude and zero dc field.}
\label{Fig:2}
\end{figure}

Fig.\ \ref{Fig:2} (a) depicts the $T$-dependence of 
the 
susceptibility  determined from the dc magnetization
of our samples
measured under both
zero-field-cooled (ZFC) and field-cooled (FC) conditions with the
field $B$ = 20\,Oe applied $\parallel$ $ab$. 
Bulk superconductivity of our samples
is
confirmed by the sharp diamagnetic signal of the ZFC data
at low $T$. Sample S2 does not show any difference between ZFC and FC curves 
above $T_c$. But for sample S1, a clear
splitting is observed below 100~K (see the inset of Fig.\ \ref{Fig:2}(a)). 
The kink in the ZFC data is attributed to the
freezing temperature of a SG-phase, $T_f \approx 60K$ at 
$B=20$~Oe. $T_f$ decreases with increasing field and at 1~T the splitting is 
observed below 15~K, only (see Fig.\ref{Fig:2}(b)). 
In addition to the ZFC-FC splitting 
an 
frequency-dependence of the ac susceptibility (see inset Fig.\ref{Fig:2}(b))
was observed for crystal S1.
\begin{figure}[b]
\includegraphics[width=20pc,clip]{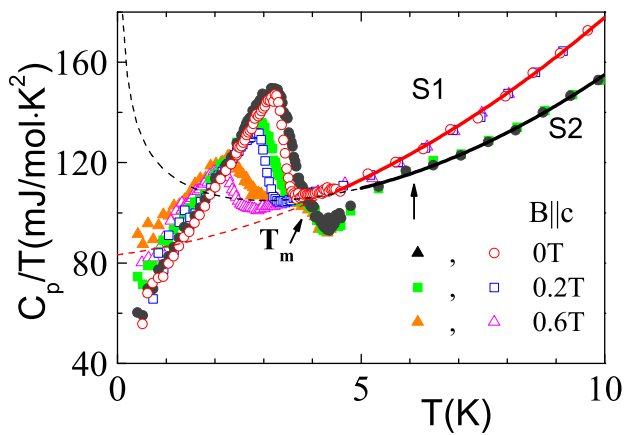}
\caption{(Color) The electronic specific heat of two K122 
samples S1 and S2 (interpretation of the fits: see text).
}
\label{Fig:3}
\end{figure}
Such a behavior is generic for a spin-glass phase
\cite{Stewart2001,Binder1986,Overhauser59}. 
In view of the INS data
\cite{Lee2011} one might suggest that the spin-glass we observed
is of helical-short range order type \cite{Ioffe85} and closely
related to the incommensurate spin fluctuation. Following
qualitatively
Ref.\ \onlinecite{Overhauser59} it is tempting to assume that few
magnetic impurities might convert locally the incommensurate SDW from an
{\it excited} state in the clean limit to a pinned SDW-type phase,
and finally to a glassy state
below $T_f \approx 60$~K. Further 
studies are desirable to improve quantitatively such a scenario
obtained originally  under simplifying assumptions.
In contrast, the analysis of the
$T$-dependence 
for the susceptibility of sample S2 measured at different fields 
doesn't 
show any magnetic transition. 
On the other hand, at $T > 30$~K the susceptibilities of 
our two crystals exhibit
similar $T$-dependencies as shown in Fig.\ \ref{Fig:2}(b) by choosing a 
proper $T$-range. We suppose that the magnetic defects 
of crystal S2 are 
less homogeneously distributed
compared to  those in sample S1. This should
be related to the
different preparation techniques used for both samples. Therefore, 
in some minor regions of crystal S2
with a sufficiently high local defect concentration, also 
magnetic 
clusters are formed with decreasing $T$. 
We suppose that these clusters are responsible for the 
flattening of $\chi(T)$
below 100~K. On the other 
hand, we suggest that 
a predominant part of
magnetic clusters leads to the
formation of a Griffiths 
(G) phase \cite{CastroNeto1998, Stewart2001}. 
Hence,
we ascribe the observed anomalous
power-law of the susceptibility at low $T < 30$~K 
to the formation of a  Griffiths-phase: 
 \begin{equation}\label{eq3}
\chi(T)= \chi_0+Cu_{\mbox{\tiny G}}/T^{1-\lambda_{\mbox{\tiny G}}} \quad ,
\end{equation}            
where $\lambda_{\mbox{\tiny G}} \approx 0.67(3)$.
At high $T$, $\chi$(T) of sample S2 
 can be fitted by a Curie law \cite{Stewart2001}:
\begin{equation}\label{eq2}
\chi(T)= \chi_0+Cu_{\mbox{\tiny 0}}/T \quad ,
\end{equation}
with the
constant
susceptibility 
$\chi_0 = 4.4\cdot 10^{-4}$emu/mol
and the Curie constant $Cu_0 = $~0.115 emu/mol$\cdot$K. 
A fit of our 
data by 
Eqs.\ (\ref{eq3},\ref{eq2}) is shown in Fig.\ \ref{Fig:2}(b) (a 
zoom for high-$T$  is shown in 
the supplement (Fig.\ref{Fig:4}ap). 
The value of $Cu_0$ provides direct insight into the defect 
concentration $\delta$, if 
a 
microscopic model is adopted.
We are enforced to adopt
such a "point"-defect model, anyway, since no other  
minority phase could  be detected so far
\cite{remminorphases}. For example, this value of $Cu_0$ can be explained by 
about $\delta_{S2}\sim 0.6$ at.\% of Fe$^{+2}$ interstitials with an 
effective 
moment 5.4$\mu_B$.
The susceptibility of sample S1 cannot be fitted by Eq.\ \ref{eq2} 
at $T<300$K. This suggests a stronger interaction between the LM which is probably 
related to a higher defect concentration $\delta_{S1}$ in sample S1 
compared to $\delta_{S2}$ if the same microscopic model is adopted for both 
crystals. Note, that we do 
not expect that $\delta_{S1}$ essentially exceeds $\delta_{S2}$ 
since both samples have similar $T_c$.   
 
We show
$C(T)_{p}/T$ for both  samples 
in Fig.\ \ref{Fig:3}. 
In case of a SG-phase the magnetic contribution
$C_{m}$ to the SH varies approximately linearly at $T<T_f$ 
similar to the usual
electronic contribution $C_{el}$ in 
case of a FL
\cite{Binder1986, Martin1980, Overhauser59}. Empirically, 
this behavior can be 
approximated by $C_{m}\approx c_1T+c_2T^2$, where $c_1$ and $c_2$
are 
constants. Then, the normal state 
SH of sample S1 can be described by: 
\begin{equation}\label{eq4}
C_p(T)=(\gamma_{el}+c_1)T+c_2T^2+\beta_3 T^3+\beta_5 T^5.
\end{equation}
On the other hand, in case of a G-phase (sample S2), 
$C(T)_G/T \propto \chi(T)$ \cite{CastroNeto1998, Stewart2001}. Hence, for
the SH we have:
 \begin{equation}\label{eq5}
C_p(T)=\gamma_{el}T+\gamma_GT^{\lambda_G}+\beta_3 T^3+\beta_5 T^5 \ ,
\end{equation}
where $\lambda_{\mbox{\tiny G}} = $~0.67 according to our magnetic 
measurements.
The data of our two crystals can be fitted by 
Eqs.\ (\ref{eq4}) and (\ref{eq5})
using the same lattice  
contribution $\beta_3=$0.68(2) mJ/mol $\cdotp$K$^4$ and 
$\beta_5=$10$^{-4}$(0.5)mJ/mol $\cdotp$K$^6$ with 
$\gamma_{el}$=60(10)mJ/mol$\cdotp$K$^2$, 
$\gamma_G$=56(10)mJ/mol$\cdotp$K$^{1+\lambda_G}$, $c_1$=23(5)mJ/mol$\cdotp$K$^2$ 
and  
$c_2$=2.6(5)mJ/mol$\cdotp$K$^3$, respectively.  
Below 6~K the experimental data for
crystal S2 deviate from the fitting 
curve (Fig.\ \ref{Fig:3}).
This behavior is
accompanied by kinks in the resistivity and the
SH data (see inset Fig.\ \ref{Fig:1}(b) and Fig.\ 3). It seems that 
$\rho(T)$ 
approaches a FL-like behavior
for $T< 6.5$~K. 
At $T_m \sim$4~K, slightly above $T_c$, (see 
Fig.\ \ref{Fig:3}) another magnetic anomaly is well visible in the
SH data. In analogy with \cite{Ubaid-Kassis2010} we attribute 
the observed behavior for 
crystal S2 to a freezing 
of the G-phase and the formation of a cluster glass (CG) phase.
Note that in 
case of metallic SG \cite{Martin1980, Martin1979}
a large magnetic contribution to SH 
is expected. 
For example, AuFe SG \cite{Martin1980} with 1 at. \% of Fe has a 
magnetic contribution $C^{Fe}_m/T$=3.3-4.6mJ/K$^2$g$\cdotp$at. at 
$T=0$.
 Assuming that 
$C_m/T$ at 
$T$=0 is nearly independent of the LM concentration 
\cite{Martin1979,Overhauser59}, 
we arrive at $c_1\sim ZC^{Fe}_m/T$=17-23~mJ/mol$\cdotp$K$^2$ 
in the case of sample S1 (with $Z$=5 atoms per f.u.) 
in accord with our findings. 
Remarkably, our findings are in a semi-quantitative
agreement with the 'universal' Overhauser
relation \cite{Overhauser59} 
(obtained in the low $T$-approximation) 
rewritten in our notation
\begin{equation}
c_1T_f=\frac{\pi^2}{9}\frac{S(S+1)}{2S+1}N_{\rm A}k_{\rm B}Z
\delta_{\rm S1},
\label{Overhauser}
\end{equation}
where $N_{\rm A}$ is the
 Avogadro's number. From Eq.\ (\ref{Overhauser})
we have
$\delta_{\rm S1}\approx 2$ at\% for
Fe$^{+2}$ ($S=2$). Then
we may
propose a realistic microscopic scenario
for excess Fe
induced spin-glasses
with incommensurate short-range
order.
Hence, 
the observed additional 
$C_m$ due to SG phase naturally
explains variations and even very 
high nominal values of $\gamma_{\rm n} {>} 90$~mJ/mol$\cdotp$K$^2$ reported 
recently 
\cite{Fukazawa2011, Kim2011, Hafiez2011}.

The presence of disordered magnetic phases inevitably 
leads to deviations from the
FLB for 
$\rho(T)$
\cite{Stewart2001}. In  case 
of a SG-phase  the  NLB shows up
for $\rho$ in the anomalous exponent $\alpha \approx 1.5$ 
\cite{Rivier1975}.
However, this value 
differs from our observed value $\alpha = 1.77$ 
for crystal S1  (\ref{Fig:1}(b)). For the
G-phase, $\alpha \leq 1.5$ is 
suggested 
by most of the 
experimental data 
\cite{Stewart2001} .
This is far from the effective value
$\alpha=1.76$ for crystal S2.
We ascribe this
puzzle 
to
multi-band effects, if 
the impurity scattering dominates only in a part of the 
Fermi surface (FS).
In fact, for the sake of
simplicity, we consider two parallel channels with different
$\rho(T)$-laws  (see also the supplement). 
As shown in Fig.\ 
\ref{Fig:1}(b), our data are well fitted,
if for one part of the Fermi surface
NFLB is assumed with 
$\rho_{m}(T)=\rho_{m0}+A_{m}T^{\alpha}$ and $\alpha = 1.5$ 
(sample S1) or $\alpha = 1$ (sample S2)
whereas for the remaining parts the 
standard FLB holds: 
$\rho_{n}(T)=\rho_{n0}+A_{n}T^{2}$.
In this case for the $T$-dependence of the effective resistivity we have:   
 \begin{equation}\label{eq1}
\rho_{eff}(T)=\rho_{m}(T)\left[1+\frac{\rho_{m}(T)}{\rho_{n}(T)}\right]^{-1} \ .
\end{equation}

Thus, the obtained value of $\gamma_{el}$=60(10)mJ/mol$\cdotp$K$^2$ 
might be considered as a new 
representative
intrinsic value for a perfectly
clean K122 system without LM. Such a value is strongly supported by the 
observation of a large residual contribution 
$\gamma_{\rm r}\approx 15$~mJ/mol$\cdot$K$^2$ 
(extrapolated to $T=0$) which amounts about 60\%
of the linear contribution $c_1$ of 
the spin-glass (SG) above $T_c$, in other words the SG is somewhat suppressed
at low $T$ where the superconductivity is most dominant \cite{Hafiez2011}. 
Moreover, independent theoretical estimates, including also a 
Kadowaki-Woods analysis, yield similar values of $\gamma_{el}$
\cite{Hafiez2011}. Following the analysis given there, we are left 
with an el-spin fluctuation coupling constant
$\lambda_{sf}\lesssim$1.
We note that similar residual contributions have been observed
also for other pnictides and chalcogenides 
but are 
often ascribed to spatial phase separation into a superconducting and a magnetic
region \cite{Stewart2011}. Such a scenario can be excluded 
for our samples which show no secondary phases.
The obtained  
value of $\gamma_{el}$ allows us to re-estimate 
$\Delta C_{el}/\gamma_{\rm el}T_c$ which amounts 
now 0.8(2) using $\Delta C_{el}/T_c\approx$50 
mJ/mol$\cdotp$K$^2$. Note, 
that this {\it new} $\Delta C_{el}/\gamma_{n}T_c$ value 
is close to the predictions for $d$-wave 
or $p$-wave pairing 
\cite{Stewart2011, Mackenzie2003}. 
Hence, to clarify the symmetry and the nodal structure
of the SC order parameter, a careful sample characterization
with respect to the presence of defect induced LM
might be important for K122 and other Fe-pnictides.
Naturally, the observed jump values doesn't fit
the strong pair-braking relation
$\Delta C_{el}\propto T^3_{c}$ \cite{Kogan2010}. 
To the best of our knowledge, K122 is a rare case
for 122 supeconductors {\it not} to fit to this "universal" relation.
We attribute the weak
pair-breaking in K122 to magnetic
intraband/or scattering between parts of FSs
with the {\it same} sign of the order parameter
provided by the SG and to the nearly absence of pair-breaking
due to nonmagnetic impurities for scattering in between gap regions of 
{\it different}
signs for intraband and interband scattering as well suggested
by the unusually low residual resistivity $\rho_0$ of our samples.
Another important fact is that in K122 with a SG 
phase no Pauli limiting for the upper critical field has been observed 
\cite{Hafiez2011} in contrast to La-1111 samples 
where LM cause a sizable Pauli limiting effect \cite{Grinenko2011}. 
This might indicate that the freezing of the SG reduces the polarization 
effect for itinerant electrons from the LM compared to a paramagnetic state.     
In general, the study of how  superconductivity 
is affected by a SG like state and vice versa 
is a challenging issue 
in the framework of the 
old problem of coexisting magnetism and superconductivity 
not studied in detail
since it has been
observed in few cases, only,
 e.g.\ (Lu,Gd)Ni$_2$B$_2$C  
\cite{bud'ko10}, 
UPt$_3$
 \cite{Stewart2001, Joynt2002}. 
(Note, that in case of Fe$_{1+y}$(Te$_{1-x}$,Se$_x$) for $y \ll 1$ and 
$x \sim $ 0.2-0.4,   
a SG phase 
in between an AFM
region and a superconducting one has been detected 
recently by neutron scattering
\cite{Naoyuki10}).
Anyhow, a better understanding of this interplay can be helpful  
for a deeper insight into
these glassy states being in the focus
of modern solid state physics including the interplay of disorder and
quantum criticality \cite{TVojta12,Stewart2001}.

To summarize,
analyzing low-$T$ transport and thermodynamic  
data we found that in K122 disordered magnetic 
phases (Griffith and spin glass-like) may occur near 
superconductivity and  even coexist with it, 
especially if not all electrons are strongly
affected by the glassy magnetic subsystem. 
Our data indicate that 
excess Fe
provides
 the corresponding local magnetic 
moments (LM).
The observed deviations from the standard Fermi-liquid 
behavior 
in the resistivity, the unusual large
magnetization, the anomalously 
large value of the nominal Sommerfeld coefficient $\gamma_{n}$ 
and the 
small value of
$\Delta C_{el}/\gamma_{n}T_c$ are ascribed to 
the coexisting 
disordered
magnetic phases. The observed deviation from 
$\Delta C_{el}\propto T_{c}^{3}$ behavior suggests 
a weak pair-breaking due to the LM.
To confirm
the proposed above
microscopic nature
of the LM, the glass-phases, and 
to determine also 
the superconducting gap symmetry 
further experimental and theoretical studies are needed.
We believe that LM 
(like considered here) might be of interest 
also for other
pnictides since LM can affect various thermodynamic 
properties both in the  
normal and in the superconducting state
as shown here and in Ref.\ \onlinecite{Grinenko2011} for the 
case of As vacancies in La-1111. 
In particular, also for LiFeAs due to the vicinity of various competing 
phases sizable effects from the presence of a very small LM concentration
is expected. This might explain e.g.\  
the rather different superconducting
properties 
even for different parts of the same single
crystal \cite{baeck11}.

\vspace{0.1cm}

We thank
A.\ Chubukov,
K.\ Kikoin, M.\ Kiselev, M.\ Vojta, 
U.\ R\"o{\ss}ler, 
D.\ Evtushinsky, 
J.\ van den Brink,
V. Zabolotnyy, and S.\ Borisenko
for 
discussions. 
Our work
was supported by the DFG (SPP 1458
and Grants No. GR3330/2 ) 
and the IFW-Pakt f\"ur Forschung.

\vspace{17cm}
\newpage
\widetext{\Large \bf EPAPS supplementary online material:
 \char`\"{} 
KFe$_{2}$As$_{2}$: coexistence of superconductivity and local moment derived
spin-glass phases"}\\

\vspace{0.15cm}
\noindent
V.\ Grinenko, M.\ Abdel -Hafiez, S.\ Aswartham, A.U.B.\ Wolter-Giraud, 
C.\ Hess, M.\ Kumar,
S.\ Wurmehl, K.\ Nenkov,\\
 G.\ Fuchs, B.\ Holzapfel, S.-L.\ Drechsler, and B.\ B\"uchner\\
{\it \small IFW-Dresden, P.O.\ Box 270116, D-01171 Dresden, Germany}
\begin{figure}[b!]
\includegraphics[width=28pc,clip]{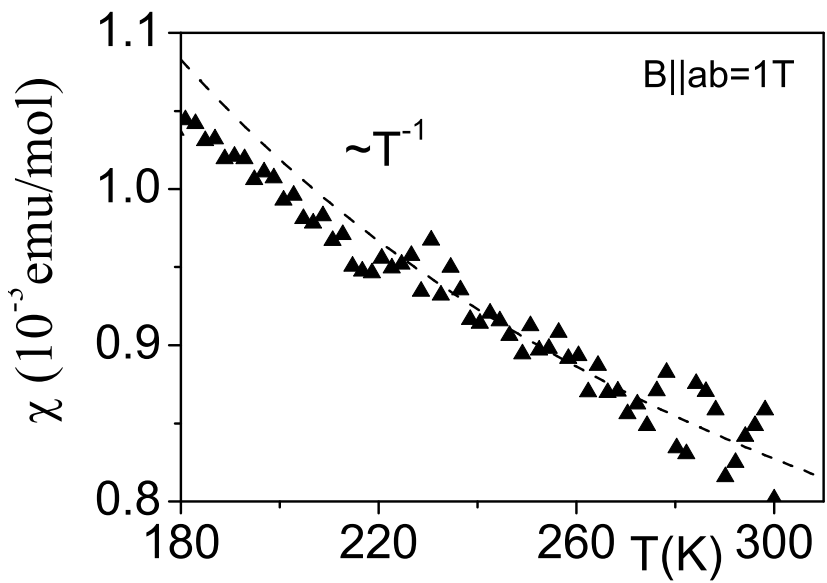}
\caption{ap (Color) 
$T$-dependence of the molar susceptibility of sample S2 at
$B$//$ab$ = 1\ and high $T$. The interval in between 220-300~K has been
used for the fit according to 
Eq.\ (\ref{eq2}) given in the main text.}
\label{Fig:4}
\end{figure}

\vspace{1cm}
\noindent
{\bf Details of the resistivity fit}\\

The Fermi surface of K122 consists 
of  4 sheets with 
comparable partial
densities of state (DOS) \cite{Hashimoto2010}. 
To specify our model, we suppose that the scattering by magnetic impurities
becomes dominant only for 
{\it one} of these 4 FS, whereas the remaining 3
are almost unaffected. This way FLB is conserved there.
Taking into account that the sample with a slightly 
higher value of RRR$_5\approx 480$ shows FLB with $\rho_{FL0} \approx 0.47 \mu
\Omega$cm and $A_{FL}=0.030(7)\mu \Omega$~cm/K$^2$ \cite{Hashimoto2010} and 
assuming nearly equal contribution of each sheets to $\rho$ we adopted 
for $\rho_{n}(T)$=4/3$\rho_{FL}(T)$. The best fit 
from 5.5 to
15~K with this $\rho_{n}$ gives for $\rho_{m0}=14.4\mu \Omega$cm, $A_{m}=0.358 
\mu \Omega$cm/K$^{1.5}$ for S1 
crystal and $\rho_{m0}=3.93 \mu \Omega$cm, $A_{m}=2.3\mu \Omega$cm/K for 
crystal S2, respectively. The essentially, higher value of $\rho_{m0}$ for S1 
is consistent with the
higher amount of magnetic defects as estimated from 
our susceptibility data.

\begin{thebibliography}{100}
\bibitem{remminorphases}In spite of $\rho$-data being less sensitive
to minor phases but described by the same magnetically disordered
models as the specific heat and the susceptibility,
we regard the presence of such phases as  unlikely.
\end{thebibliography}
\end{document}